\documentclass[preprint,aps]{revtex4}
\usepackage{graphicx} 
\usepackage{amsmath,amsfonts,amssymb}

\begin{document}

\title{Interplay between two mechanisms of resistivity}
\author{A Kapustin and G Falkovich}
\date{\today}
\begin{abstract}
Mechanisms of resistivity can be divided into two basic classes: one is dissipative (like scattering on phonons) and another is quasi-elastic (like scattering on static impurities). 
They are often treated by the empirical Matthiessen rule, which says that total resistivity is just the sum of these two contributions, which are computed separately. This is quite misleading for two reasons. First, the two mechanisms are generally correlated. Second, computing the elastic resistivity alone masks the fundamental fact that the linear-response approximation has a vanishing validity interval at vanishing dissipation. Limits of zero electric field and zero dissipation do not commute for the simple reason that one needs to absorb the Joule heat quadratic in the applied field. Here, we present a simple model that illustrates these two points. The model also illuminates the role of variational principles for non-equilibrium steady states.

\end{abstract}
\maketitle
\section{Introduction}
This is a methodological note intended to explain the basic interplay  between the two mechanisms of resistivity using a simple model analyzed in more detail in \cite{AK1}.

Consider a classical particle that moves under the influence of a uniform force $E$ in a medium with the temperature $T$ and randomly placed elastic scatterers. We denote the momentum relaxation rate due to elastic scattering (averaged over all momenta) by $\rho_e$. Finite temperature provides for additional momentum relaxation with the rate $\nu$ and for a random force, which leads to diffusion in the momentum space with the diffusivity $\nu T$. 

We define the resistivity $\rho$ as a linear-response factor relating the mean momentum to the force: $\bar p=E/\rho$. It is, thus, the mean relaxation rate of the momentum direction.  An empirical Matthiessen rule (M-rule) suggests that (see, e.g. \cite{AM})
\begin{equation}\rho\approx \nu+\rho_e\,.\label{Mat}\end{equation}
Just how off is the estimate \eqref{Mat}? Below, we show that the rule is exact when the elastic scattering is momentum-independent in two dimensions. We also compute $\rho$ in two limits where one or another mechanism dominates and show that the addition of another mechanism generally enhances resistivity much more than the  Matthiessen rule suggests.

It is also clear that computing the elastic contribution in the limit $\nu\to0$ does not make much sense since energy conservation requires
\begin{equation}\nu\langle p^2\rangle=E\cdot\bar p=E^2/\rho\,.\label{EC}\end{equation}
That means that the $E^2$-corrections to the linear-response theory diverge as $1/\nu$; that is, nonzero friction is necessary for the elastic resistivity to make sense. The behavior of the nonequilibrium distribution function in this limit iss discussed in detail in \cite{AK1}. On the contrary, we show below that the validity interval of the linear-response theory expands as $1/\rho_e$ when $\rho_e\to0$.

\section{The basic equation}
Let us consider the simplest kinetic (Fokker-Planck) equation on the momentum distribution $f({\bf p},t)$, which satisfies
the equation:
\begin{equation}\label{eq:W0}
\frac{\partial f}{\partial t}=\frac{\partial}{\partial p^i} (\nu p_i-E_i)f+T\nu\Delta f+\hat I f\ .\end{equation}
The first term on the right is due to a constant force and linear friction. Random kicks from the thermostat provide diffusion in the momentum space, described by the Laplacian. Let us stress that the $\nu$-terms is the simplest model of a thermostat; our classical consideration only qualitatively corresponds to the scattering of quantum electrons on phonons. On the other hand, it describes a variety of classical stochastic systems. The last term is a linear operator $\hat I$ describing elastic scattering.

Statistical isotropy of scattering means that the angular harmonics are eigenfunctions of the scattering operator: $\hat If_l=\gamma_lf_l$. Mean momentum and resistivity are determined by the first-harmonic rate, which we denote $\gamma_1=\gamma$. It generally depends on $p$. One universal limit is that of scattering by a small angle proportional to the time of interaction and inversely proportional to $p$. For a finite-range scattering potential, the time is also $\propto 1/p$ so that the deflection angle is $1/p^2$.  The rate of meeting scatterers is proportional to the momentum $p$. As a result, small-angle scattering leads to angular diffusion: $\hat I={W\over p^3}\Delta_\Omega$ so that $\gamma=(d-1)Wp^{-3}$  \cite{vH}. That typically occurs
when the average momentum exceeds the potential strength measured by $W$  (proportional to the 2-point correlation function of the potential \cite{AK1}). For lower momenta, the scattering is by angles of order unity; the rate of momentum loss in many such cases is proportional to the momentum itself or independent of momentum.  It is also instructive to consider power-law functions $\gamma(p)\propto p^a$ with different $a$.  

For every momentum, two mechanisms have their scattering rates added 
according to \eqref{eq:W0}. The first-harmonic correction to $f({\bf p})$ is inversely proportional to the sum of the rates for every momentum $p$. The conductivity is proportional to the correction integrated over momenta (see e.g. \cite{AM}). The total resistivity (inverse conductivity) is then bounded by $\rho\geq \nu+\rho_e$, where $\rho_e=\gamma$ is the average of $\gamma(p)$ over $p$.  
Only when both relaxation rates are independent of momenta, we have the M-rule equality: $\rho= \nu+\rho_e$. Here we shall see that in our case, it is enough that elastic scattering is momentum-independent for the Matthiessen rule to hold.

The model is characterized by two dimensionless parameters, $B=\gamma/\nu$ and $F=E/\nu T^{1/2}$. The first one determines the relative role of the two mechanisms of resistivity. The second one characterizes the strength of the field.

Without the external force, $E=0$, the equation has an equilibrium Maxwell isotropic solution $f_0(p)\propto \exp(-p^2/2T)$, which is independent of $\nu$. It realizes the maximum of entropy $S=-\int f\log f\,d{\bf p}$ for a given mean energy $\int (p^2/2)f\,d{\bf p}$.

Without scattering, $\hat I=0$, the solution has a Gaussian form for arbitrary $E$ since \eqref{eq:W0} has a symmetry which shifts ${\bf p}$ and ${\bf E}$ simultaneously:  \begin{equation}
 f_0(p,\theta)=(2\pi)^{-d/2}\exp[-|{\bf p}-{\bf E}/\nu|^2/2T]\,.\label{Max}   
\end{equation}  
The distribution \eqref{Max} gives a linear current-field relation, $\langle p\rangle=E/\nu$, for any $E$. It also realizes the entropy maximum under the condition of the energy production-dissipation balance, $\nu\langle p^2/2\rangle={\bf E}\langle {\bf p}\rangle$. Indeed, it realizes the extremum of the functional $\int f[-\log f+\lambda(\nu p^2/2-{\bf p}{\bf E})]\, d{\bf p}$. 
Even though \eqref{Max} looks like a shifted equilibrium whose entropy does not depend on $E$, it is a non-equilibrium state with energy dissipation and entropy production.

\section{Linear response at weak elastic scattering}
Let us describe the effect of scattering on the distribution and the linear resistivity in the limit $B=\gamma/\nu\to0$. Even though \eqref{Max} is valid at arbitrary $E$, we could compute the $\hat I$-corrections to it only in the limit  $F=E/\nu T^{1/2}\to 0$. We assume $f=f_0(p)+f_1({\bf p},E)+f_2({\bf p},E,\gamma)$, where $f_0 \propto e^{-p^2/2T}$ and $f_1=f_0(pE/\nu T)\cos\theta$ are given by  \eqref{Max}.  We assume $f_1\gg f_2\propto BF\propto \gamma E$.  Substituting it into \eqref{eq:W0} gives 
\begin{equation}\label{eq:W1}
 \frac{\partial}{\partial p^i} E_if_2-\frac{\partial}{\partial p^i}\nu p_i  f_2-T\nu{1\over p}\frac{\partial }{\partial p }p\frac{\partial f_2}{\partial p }-{T\nu\over p^2}\Delta_\Omega f_2 \approx \hat If_1=\cos\theta\gamma(p)f_0{pE\over \nu T}\ .
\end{equation}
Since we are interested in the contribution of scattering to resistivity, we consider only the first angular harmonic, $f_2=f_0\chi(p)\cos\theta$, which satisfies the equation
\begin{equation}
p{\partial \chi\over\partial p}-{T\over p}{\partial \chi\over\partial p}-T{\partial^2 \chi\over\partial p^2}+{T(d-1)\chi\over p^2}=-\gamma(p){pE\over\nu^2T}\,.\label{gen}
\end{equation}
For a general $\gamma(p)\propto p^a$, the solution has different asymptotics for large and small $p$: $\chi\propto -p^{a+1}E/(a+1)\nu^2 T$ for $p\gg T$ and  $\chi\propto-Wp^{a+3}E/[d-1-(a+3)^2]\nu^2 T$ for $p\ll T$. In this case, one needs numerics to compute the solution and the correction to resistivity from weak elastic scattering. Fortunately, for two specific (and physical!) values of $a$, the solution has a simple power form, $\chi=bp^{c}$, where $b,c$ are constants to be determined. Only solutions with $b<0$ make physical sense since the scattering must diminish the current. 
Action by the first term in \eqref{gen}  gives $-b\nu cp^c\cos\theta$. 
Action by the third term gives $\,b\nu cp^{c-2}[2p^2-cT]\cos\theta$.
Action by the fourth term gives $ b(d-1)\nu T p^{c-2}\cos\theta$. Then the equation  \eqref{gen} gives
\begin{align}
b\nu cp^{c}+b\nu p^{c-2}T \left[d-1- c^2\right] = -\gamma(p){pE\over \nu T}\,.\label{f2}
\end{align}
Since the terms on the left have different powers of $p$, the power-law solution exists only for $c=0$ and $c=\sqrt{d-1}$. The former case corresponds to the small-angle scattering when $\hat I =Wp^{-3}\Delta_\Omega$ and $\gamma(p)=(d-1)W p^{-3}$. 
In that case we get
\begin{equation}
    f_2=-{ W E \over (\nu T)^2}f_0\cos\theta\,.\label{b}
\end{equation}
The correction \eqref{b} diminishes the current and gives corrections to the conductivity and  resistivity: 
\begin{equation}
  \sigma={1\over\nu}\left(1 -{ W \over  \nu T^{3/2}}\right)\,, \quad
   \rho \approx\nu  +{ W \over   T^{3/2}}\,. \label{sig}
\end{equation} 
In the next section, we compute the elastic resistivity for small-angle scattering: $\rho_e= {(d-1)W\sqrt {2\pi}}/{32 T^{3/2}}$. Comparison with that value shows that the 
Matthiessen's rule, $\rho= \nu+\rho_e$, is quite off in the limit of weak elastic scattering. For $d=3$ the $W$-addition to the resistivity (\ref{sig}) is more than six times larger (and for $d=2$ twelve times larger) than the rule predicts since there is a strong positive correlation between the two mechanisms of the momentum relaxation. Indeed, an angular scattering enhances the frictional relaxation of the $x$-momentum by bringing more particles from other directions. 

The case $c=\sqrt{d-1}$ for $d=2$ corresponds to  $\gamma$ independent of $p$: \begin{equation}     f_2=-{\gamma p E  \over \nu^2 T}f_0(p)\cos\theta=-f_1{\gamma\over\nu}\,,\label{f12} \end{equation} which gives the conductivity correction $-\gamma /\nu^2$ and the  resistivity as follows: 
\begin{equation}\rho=\nu +\gamma \,.\label{res1}\end{equation}
We see that only for the momentum-independent scattering rate in two dimensions is the Matthiessen's rule valid. In this case, \eqref{res1} is valid for arbitrary relation between $\gamma$ and $\nu$. Indeed, the linear-response first angular harmonic is exactly equal to 
\begin{equation}
    f_1={pE\over (\nu+\gamma)T}f_0\cos\theta\,.\label{nugam}
\end{equation}

The terms we neglected are quadratic in $E$ and contribute to the zeroth and second harmonics. They can be accounted for in the next orders. The terms cubic in $E$ contribute to the third angular harmonic. 

\section{Low-friction limit}
One may assume that in the limit $\nu\to0$, $T\to0$ one can neglect the thermostat-related terms and write:
\begin{equation}\label{eq:nu0}
\frac{\partial f}{\partial t} +\frac{\partial}{\partial p^i} E_if={\hat I} f\ .
\end{equation}
Yet this equation does not have a steady state for the simple reason that the second term pumps energy while the last one does not change it. Despite that, one can find the resistivity within the linear response theory, assuming that $f({\bf p})=f_0(p)+ f_1({\bf p})+O(E^2)$, where $f_0(p)$ is isotropic and $f_1\propto E$. Then for small-angle scattering we can write 
\begin{equation}\label{eq:nu0}
 \frac{\partial}{\partial p^i} E_if_0={W\over p^3}\Delta_\Omega f_1\ .
\end{equation}
\begin{equation}\label{eq:nu1}
f_1=-E\cos\theta{p^3f_0'\over (d-1)W}
\end{equation}
Taking $f_0=\delta(p-p_0)$, we obtain ${\bar p}=\frac{2p_0^3E}{(d-1)W}$, which gives the resistivity for a given energy $E_0=p_0^2/2$: 
\begin{equation}
\rho_e(E_0)={(d-1)W\over 2(2E_0)^{3/2}}\,.\label{rho0}      
\end{equation}
Resistivity at a fixed temperature is obtained by using $f_0\propto \exp(-p^2/2T)$:
\begin{equation}
\rho_e(T)={\sqrt{2\pi}(d-1)W\over 32T^{3/2}}\,.\label{rho1}   
\end{equation}

Let us now account for small friction. The Matthiessen rule would predict just adding $\nu$ to resistivity: $\rho={\sqrt{2\pi}(d-1)W/32T^{3/2}}+\nu $. Let us show that this is not the case. We look for the correction in the form  $f=f_0(p)+f_1({\bf p},W)+f_2({\bf p},W,\nu)$, where $f_0 \propto e^{-p^2/2T}$ and $f_1=Ef_0\cos\theta{p^4/ T(d-1)W}$ due to  \eqref{eq:nu1}.  Substituting into \eqref{eq:W0} and assuming  $f_1\gg f_2$, we obtain (for $d=3$):
\begin{equation}
f_2=f_1{\nu (9pT- 2p^3)\over W}\,.\label{nu1}
\end{equation}
Non-surprisingly, friction decreases the number of fast particles and increases the number of slow ones. 
The resulting resistivity is as follows:
\begin{equation}
\rho={\sqrt{2\pi}W\over 16T^{3/2}}+\nu {630\pi\over512} \,.\label{rho2} \end{equation}

For a general case of scattering by order-unity angles, we can simply put $\hat I f_1=-\gamma f_1$, which gives $f_1=f_0 E\cos\theta/\gamma $. The elastic resistivity is simply $\gamma $
and $\rho=\nu+\gamma$.

We thank A. Finkelstein and B. Spivak for useful discussions. The work of A. K. was supported in part by the U.S.\ Department of Energy, Office of Science, Office of High Energy Physics, under Award Number DE-SC0011632 and by the Simons Investigator Award. The work of G.F. is supported by the Excellence Center at WIS, by the Simons grants 662962 and 617006,  the NSF-BSF grant 2020765, and by the EU Horizon grants No 873028 and 823937.

\end{document}